\documentclass[showpacs,amsmath,amssymb,pre]{revtex4}

\usepackage{dcolumn}
\usepackage{graphicx}
\usepackage{amsfonts,graphicx,epsf,amsmath,amsbsy}
\usepackage{chngpage}
\usepackage[english]{babel}
\usepackage{amsmath}
\usepackage{fontenc}
\usepackage{amssymb}
\usepackage[small]{caption}
\usepackage{bm}
\usepackage{graphicx,epsf,subfigure}
\usepackage{subfig}
\numberwithin{equation}{section}
\usepackage{chngpage}

\usepackage{lipsum}

\usepackage{booktabs}

\newcommand{\be}{\begin{equation}}
\newcommand{\ee}{\end{equation}}

\begin{document}

\title{Front speed in reactive compressible stirred media}

\author{Federico Bianco}\affiliation{Institut D'Alembert University Pierre et Marie Curie, 4, place jussieu 75252 Paris Cedex 05}
\affiliation{CNRS UMR 7190, 4, place jussieu 75252 Paris Cedex 05}
\affiliation{Dipartimento di Fisica, Universit\`a ``La Sapienza'', Piazzale Aldo Moro 2, I-00185 Roma, Italy}
\author{Sergio Chibbaro}\affiliation{Institut D'Alembert University Pierre et Marie Curie, 4, place jussieu 75252 Paris Cedex 05}
\affiliation{CNRS UMR 7190, 4, place jussieu 75252 Paris Cedex 05}
\author{Davide Vergni}\affiliation{Istituto Applicazioni Calcolo, CNR, V.le Manzoni 30, 00185, Rome, Italy}
\author{Angelo Vulpiani}\affiliation{Dipartimento di Fisica, Universit\`a ``La Sapienza'' and ISC-CNR, Piazzale Aldo Moro 2, I-00185 Roma, Italy}

\date{\today}

\begin{abstract}
We  investigated a nonlinear
advection-diffusion-reaction equation for a passive scalar field. 
The purpose is to  understand how the compressibility can affect the front
dynamics and the bulk burning rate. We study two classes of flows:
periodic shear flow and cellular flow both in the case of 
fast advection regime, analysing the system at varying the
extent of compressibility and the reaction rate.  
We find that the bulk burning rate $v_f$ in a shear flow increases
with compressibility intensity, $\epsilon$, following a relation
$\Delta v_f\sim \epsilon^2$. Furthermore, the faster the reaction the more 
important the difference with respect to the laminar case.
 The effect has
been quantitatively measured and it turns out to be generally little.
For the
cellular flow, the two extreme cases have been investigated, with the
whole perturbation situated either in the centre of the vortex or in
the periphery.  The dependence in this case does not show a monotonic
scaling with different behaviour in the two cases. The enhancing
remains modest and always less than $20\%$.
\end{abstract}

\maketitle

\section{Introduction}
The dynamics of reacting species presents several issues of great
interest from a theoretical point of
view~\cite{xin,Constantin:2000p15663,Ver_12}. { Moreover it} is also 
a problem of wide application in many fields, from front propagation in
gases~\cite{combustion}, chemical reaction in liquids~\cite{chem,e95}
and ecological dynamics of biological systems (e.g. plankton in
oceans)~\cite{bio,alplmb00,guasto2012fluid,korolev2010genetic,d2010fluid}.

In the most simplest model of reaction dynamics,
the state of the system is described by a single scalar field
$\theta({\bf r},t)$, that represents the concentration of
products. The field $\theta$ vanishes in the regions filled with fresh
material (the unstable phase), equals unity where only inert products
are left (the stable phase) and takes intermediate values wherever
reactants and products coexist, i.e., in the region where production
takes place.  In their seminal contributions, Fisher, Kolmogorov,
Petrovskii and Piskunov~\cite{FKPP,f37} (FKPP) considered the simplest case
of pure reaction/diffusion and proposed the so-called FKPP model
\begin{equation}
   \partial_t \theta  =
   D \Delta \theta + f(\theta)\,,
   \label{eq:rd}
\end{equation}
where $D$ is the molecular diffusivity and $f(\theta)$ describes the
reaction process that obviously depends on the phenomenon 
under investigation. In this work, as in the original works of FKPP, 
we focus on pulled reaction, e.g. the autocatalytic reaction 
$f=\alpha\theta(1-\theta)$, where $\alpha$ is the reaction rate
and its inverse, $\tau=1/\alpha$, is the reaction time.

However, most natural phenomena take place in deformable media like fluids and therefore transport properties cannot be ignored.
If the medium is stirred, e.g. an eulerian velocity field $\mathbf{u}(\mathbf{x})$ is present, Eq. (\ref{eq:rd}) can be generalized in the incompressible case to
\begin{equation}
   \partial_t \theta + (\mathbf{u} \cdot
\mathbf{\nabla})\theta=D \Delta \theta + f(\theta)\,\,.
   \label{eq:evol}
\end{equation}
The complete mathematical description of these
phenomena is given by partial differential equations (PDE) for the coupled evolution of the velocity field and of the concentration
of the reacting species~\cite{combustion}.
Therefore the above Eq.~(\ref{eq:evol}) should be coupled with 
Navier-Stokes equations (usually in a non trivial way).
This is the general framework
for treating engineering combustion problems in gases~\cite{Peters,prudhomme,poinsot}.
In some cases, e.g.~\cite{vladimirova2003model}, the coupling can be simplified using a Boussinesq term.\\
In this work, as a further simplification, we assume that the
reactants do not influence the velocity field which evolves independently.
In such a limit the dynamics is still non trivial and it is completely
described by the above ARD Eq.~(\ref{eq:evol}) together with the proper definition of a given velocity field, $\mathbf{u}(\mathbf{x})$.
This equation has been intensively studied in incompressible media~\cite{audoly,vladimirova2003flame,acvv1,ctvv}.
In particular, it has been investigated the dependence of the front speed as a function of $D,\alpha$ and the velocity field $\mathbf{u}(\mathbf{x})$~\cite{acvv}.

On the contrary, in the case of compressible flows, the ARD problem did not receive too much attention but only recently in a mathematical framework~\cite{constantin2008propagation,benzi}.
To account for compressible flows is indeed not simple but it is a relevant issue
in combustion~\cite{Peters,poinsot}, plankton dynamics in turbulent flows~\cite{lewis2000planktonic} and also in particle-laden flows, where the particle phase can be highly compressible even in incompressible flows, because of inertia~\cite{Min_01,falkovich2001particles,bec_07}.
While the passive scalar approximation for reactive species is hardly tenable in gas combustion phenomena, it may be considered appropriate in aqueous or liquid reactions (notably plankton in oceans) and for dilute particle-laden flows.
In those cases, it may give some relevant insights for front propagation and can be used as a model for the flame tracking in some limits\cite{vladimirova2006flame}.
 
Our aim is to investigate the effects of the compressibility to the bulk burning rate
of the reaction process by studying the following PDE:
\begin{equation}
\rho \left[ \frac{\partial \theta}{\partial t}  + u_i \frac{\partial \theta}{\partial x_i}   \right] =   D_0 \frac{\partial^2 \theta}{\partial x_i^2}  +  f(\theta) 
\label{scalar}
\end{equation}
The scalar field $\theta$ represents the mass fraction of a single species of a binary mixture, while $u_i$ is the 
$i^{th}$ component of a given compressible flow, $D_0=\rho D$ is the diffusion coefficient (supposed to be constant), 
$ f(\theta) = \dot \omega$ the rate of production of the chosen species and $\rho$ the non constant density of the fluid.\\

The paper is organized as follows:
Section~\ref{sec:2} is devoted to the presentation of the model and the principal aspects of the numerical computations. In Section~\ref{sec:3} we discuss the results for the front propagation in compressible shear flows. Section~\ref{sec:4} is devoted to the case of compressible cellular flows. Finally, in Section~\ref{sec:5}, the reader can find the conclusions.

\section{The model\label{sec:2}}
The PDE model described by Eq.~(\ref{scalar}) can be derived from the equation of conservation of species, that is relevant for combustion dynamics~\cite{combustion,prudhomme}.
Let us consider two species (namely A,B) which diffuse and react together while they are
passively transported  by a compressible flow, being $\rho_A(x,y,t)$ the mass of species $A$ per unit volume, 
the conservation of species $A$ gives:
\begin{equation}
\frac{\partial{\rho_A}}{\partial t} + \frac{\partial}{\partial x_i} \left[{\rho_A (u_i+U_{A,i})} \right]=\dot \omega_A
\end{equation}
where $u_i$ is the $i^{th}$ component of the advective flow field, $U_{A,i}$ 
is the velocity of diffusion of species $A$ and $\dot \omega_A$ is the rate of production. 

Define the mass fraction $Y_k=\rho_k/\rho$,  where $\rho$ is the density of the mixture and $k=A,B$.
The species conservation can be written in terms of mass fraction as follows:
\begin{equation}
\frac{\partial{(\rho Y_k)}}{\partial t} + \frac{\partial}{\partial x_i} \left[{\rho Y_k (u_i+U_{k,i})} \right]=\dot \omega_k
\label{generale}
\end{equation}
Where $Y_A+Y_B=1$.
Moreover, if Fick's law is considered, the diffusion velocities can be defined as follow:
\begin{equation}
Y_A U_{A,i} = -Y_B U_{B,i}= - D  \frac{\partial Y_A}{\partial x_i}
\end{equation}
In the following, we assume an auto-catalytic irreversible law $A+B \xrightarrow{} 2A$:
\begin{equation}
\dot \omega_A= \alpha {\rho_A \rho_B}=\alpha \rho^2 Y_A Y_B=\alpha \rho^2 Y_A (1-Y_A)
\end{equation}
where the constant $\alpha$ controls the speed of reaction and by definition $\dot \omega_A=-\dot \omega_B.$

Thus, the evolution of the mass fraction of  species A is completely described by the following PDE:
\begin{equation}
\rho \left[  \frac{\partial \theta}{\partial t}  + u_i \frac{\partial \theta}{\partial x_i} \right]=  D_0 \frac{\partial^2 \theta}{\partial x_i^2}  + \alpha \rho^2 \theta (1-\theta )\,,  
\label{scalar2}
\end{equation}
that holds if we neglect the coupling between conservation of species equation and the conservation of energy equation.
 That is the case in which the energy released by the 
reaction is negligible and thus the momentum and energy equations evolve independently.
The left hand side of Eq.~(\ref{scalar2}) is written in non-conservative form using the continuity equation of the mixture and 
the product $\rho D=D_0$ is assumed constant (which is quite a reasonable hypothesis~\cite{Peters,prudhomme}).

Since we are interested, in the front propagation, we consider the following geometry:
\begin{equation}
 -\infty<x<\infty~~,~~0 \le y \le L         
\end{equation}
Only for the sake of simplicity we assume periodic boundary conditions in the $y-$direction and $\theta(-\infty,y,t)=1$ (burned material in a combustion terminology) and $\theta(\infty,y,t)=0$ (fresh material).

At $t=0$ the initial condition is given by:
\begin{equation}
\theta(x,y,t) = \left\{ \begin{array}{ll}
1 & \textrm{if $x<0$}\\
0 & \textrm{if $x \geq 0$}
\end{array} \right.
\end{equation}
Of course different boundary and initial conditions may be interesting.
For instance, if one is  interested in quenching issues,
appropriate initial conditions would pose $\theta$ initially localized in a region of size $\ell$
\begin{eqnarray}
\theta(x,y,0) &=& \left\{ \begin{array}{ll}
1 & \textrm{if $-\ell/2\leq x\leq \ell/2$}\\
0 & \textrm{if $x<-\ell/2$ or $x>\ell/2$}
\end{array} \right.\,.
\end{eqnarray}

Equation~(\ref{scalar2}) has been solved using a eighth-order central finite difference 
scheme in space and a fourth-order Runge-Kutta integration in time. The grid size is sufficiently
small to guarantee a good representation of the shear across the reacting region and convergence of
solutions has been verified. To compute accurately the asymptotic mean bulk burning
 rate, very long integration periods are required. 
The grid is remapped following the reacting front and the computational domain is extended
upstream and downstream from the reactive zone so that the boundary effects are negligible.

\section{Compressible shear flow\label{sec:3}}
\begin{figure}[ht!]
\includegraphics[angle=0,scale=0.5,keepaspectratio=true]{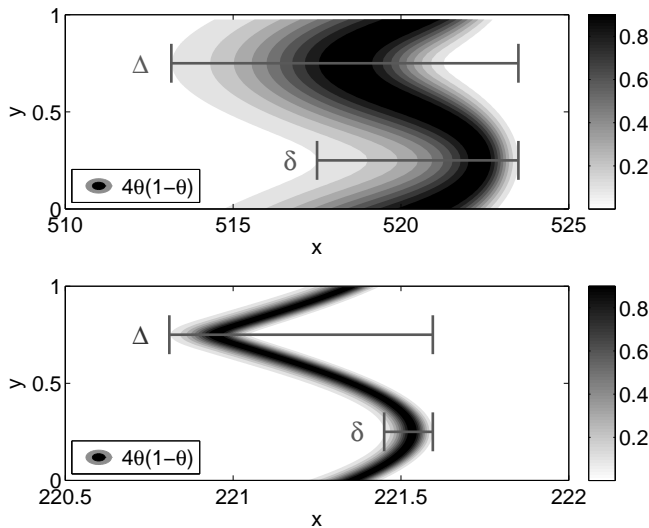}
\caption{Shape of the active part of the front (we use the function $4\theta(1-\theta)$, which is maximal for $\theta=0.5$), for a fixed Peclet number ($Pe=100$) and for two 
different reaction rates. Taking as reference the incompressible test case ($\epsilon=0$) we can define two different thicknesses. The bare front thickness as $\delta \sim \sqrt{D/\alpha}$
and the distance between the tip and the tail of the reacting region $\Delta$.
For a slow reaction (upper panel, $Da=1$), we have approximately $\Delta \approx 10$ and $\delta \approx 6$. 
For fast reaction (lower panel, $Da=100$) $\Delta \approx 0.8$ and $\delta \approx 0.15$. }
\label{spessori}
\end{figure} 
We investigate the effects of compressibility in a 2D steady-state shear flow using { the velocity field}
\begin{equation}
\bar{u}(x,y) = \left( \frac{ U_0  \sin \left(\frac{2 \pi y}{L}\right) }{ 1+\epsilon \sin \left( \frac{2 \pi x}{\lambda}\right) } , 0 \right)\,.
\label{eq:kolm}
\end{equation}
Such a choice corresponds to a Kolmogorov flow with amplitude $U_0$ and wavelength $L$, perturbed by a steady wave of wavelength $\lambda$ and magnitude $\epsilon$ accounting for the compressibility of the flow. Let us stress that the perturbation is oriented along the direction of propagation of the reactive front, i.e. the $x$ axis.\\
In order to satisfy the continuity equation, $\partial_i(\rho u_i)=0$,
it is necessary to impose a spatially dependence on $\rho$, as
$$
 \rho(x) = \rho_0 \left[ 1+\epsilon \sin \left( \frac{2 \pi x}{\lambda}\right) \right] \,.
$$
Finally, Eq.~(\ref{scalar2}) can be written in non-dimensional form:
\begin{equation}
  \frac{\partial \theta}{\partial t^*}  + u^*_i \frac{\partial \theta}{\partial x^*_i} =  \frac{1}{\rho^* Pe } \frac{\partial^2 \theta}{\partial {x^*_i}^2}  +  {\rho^*} Da \theta (1-\theta )  
\label{adimensionale}
\end{equation}
 if we define $\rho^*=\rho/\rho_0$, $x_i^*=x_i/L$, $u_i^*=u_i/U_0$, $t^*=(tU_0)/L$. \\
The adimensional { parameters} $Pe=(\rho_0 U_0 L)/D_0$ and $Da=(L \alpha \rho_0)/U_0$ are
 the Peclet and the Damk\"{o}hler numbers which define the ratio between the diffusive and 
advective time scale and the ratio between the advective and reactive time scale respectively.

In the following, we will drop the star notation and we will solve Eq.~(\ref{adimensionale}) focusing on regimes
at high Peclet number $Pe\gg1$.
Varying the Damk\"{o}hler number in a range of $Da \in [1,1000]$,  we will quantify 
the effects of $\lambda$ and $\epsilon$ on the asymptotic value of the bulk burning rate. 

The instantaneous bulk burning rate is:
\begin{equation}
v_f(t)=\int_0^{1} \int_{0}^{+\infty} \dot \omega~ dx dy=\int_0^{1} \int_{0}^{+\infty} Da \rho^2 \theta (1-\theta )~ dx dy\,,
\label{vistantaneo}
\end{equation}
while the mean or asymptotic bulk burning rate is defined as the time average of 
$v_f(t)$ over a sufficiently long interval:
\begin{equation}
v=\frac{1}{T}\int_0^{T} v_f(t) dt
\label{vmedio}
\end{equation}

To shed some light on the role played by $\lambda$ we first run a simulation in absence of compressibility
 for two different Damk\"{o}hler (slow and fast reaction) and for a fixed Peclet.
We characterize the thicknesses of the reactive front $\Delta$ and $\delta$ (see Figure \ref{spessori} for definition).  From this figure, it is clear that the faster the reaction the thinner the flame. 

\begin{figure}[ht!]
\includegraphics[angle=0,scale=0.5,keepaspectratio=true]{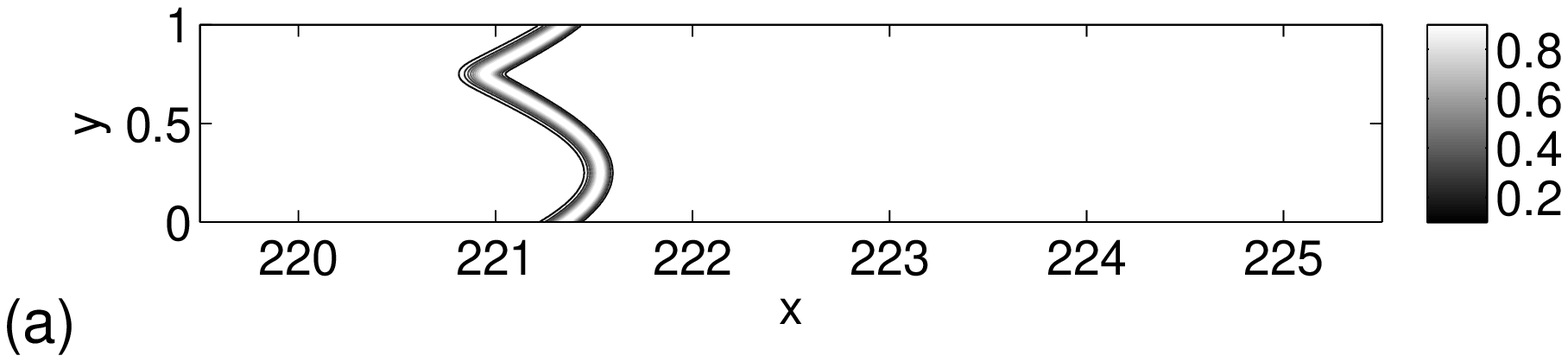} \\
\includegraphics[angle=0,scale=0.5,keepaspectratio=true]{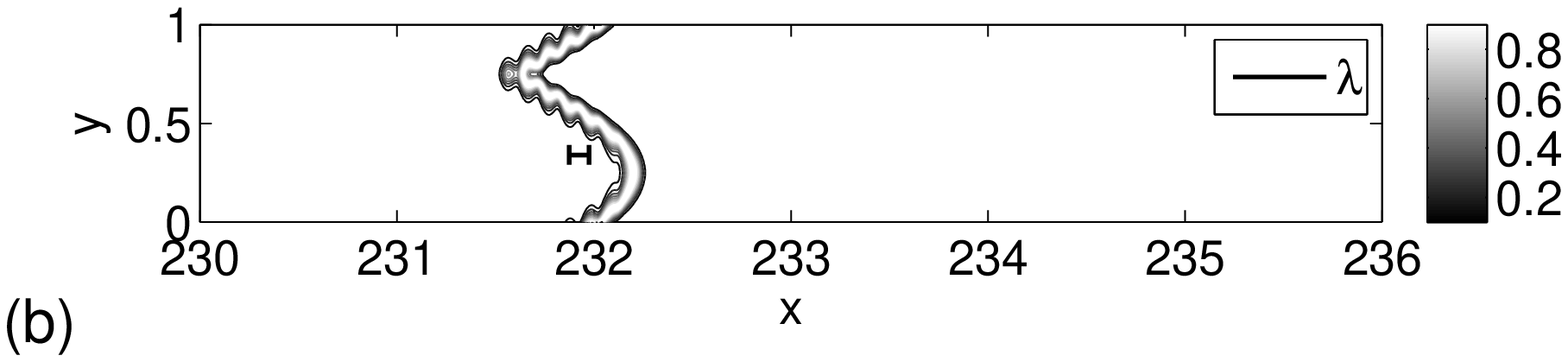} \\
\includegraphics[angle=0,scale=0.5,keepaspectratio=true]{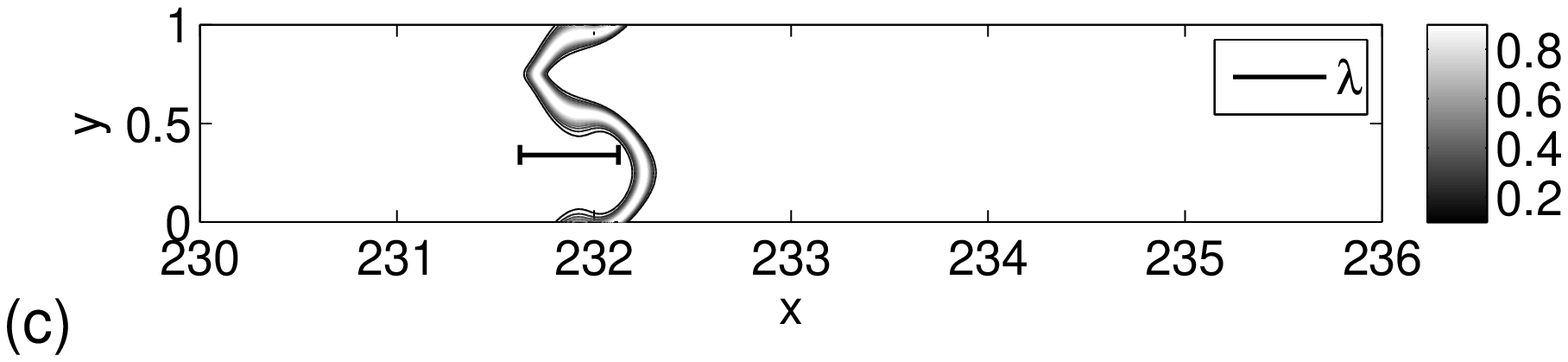} \\
\includegraphics[angle=0,scale=0.5,keepaspectratio=true]{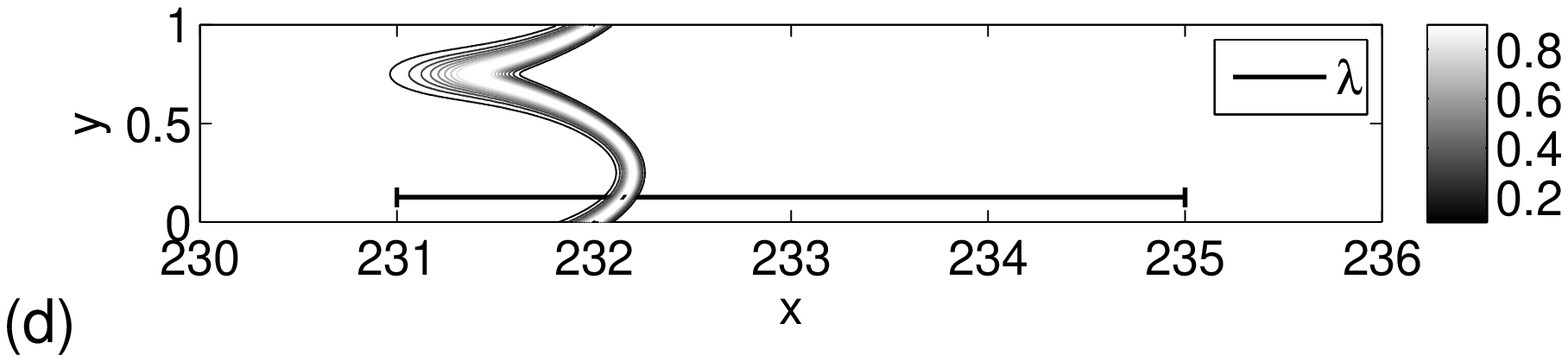} \\
\caption{Snapshot of $4\theta(1-\theta)$ for a fixed Peclet number ($Pe=100$) and
for Damk\"{o}hler number $Da=100$.
 Panel (a) refers to a incompressible simulation ($\epsilon=0$) while for the others $\epsilon=0.5$. For the compressible
tests the characteristic length $\lambda$ is set to be approximately lower (panel (b), $\lambda=0.1$),
 between (panel (c), $\lambda=0.5$) or greater (panel (d), $\lambda=4$) than the two thicknesses $\Delta$ and $\delta$.  }
\label{fronti}
\end{figure}

\begin{figure}[ht!]
\includegraphics[angle=0,scale=0.4,keepaspectratio=true]{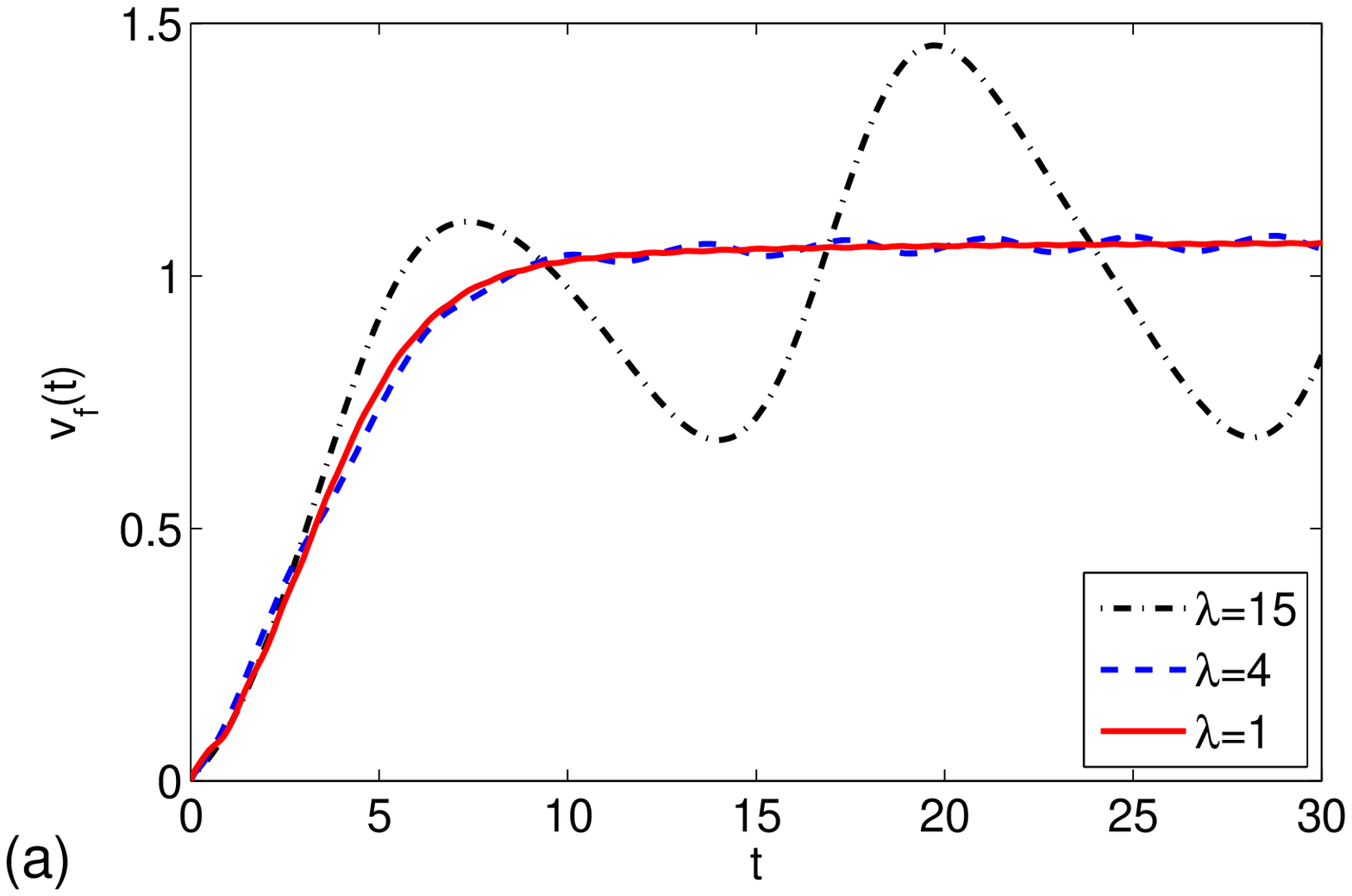} 
\includegraphics[angle=0,scale=0.4,keepaspectratio=true]{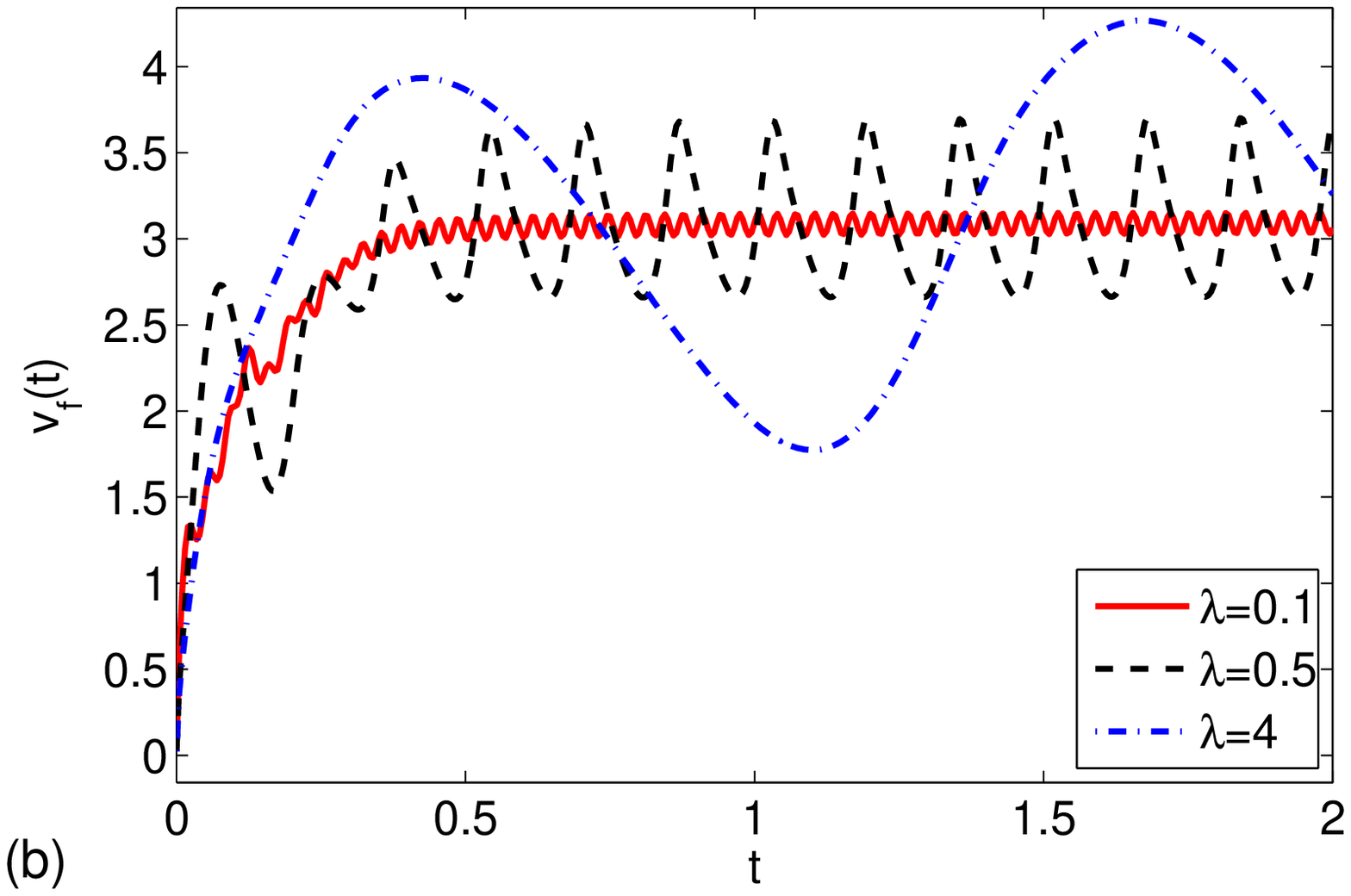} \\
\includegraphics[angle=0,scale=0.4,keepaspectratio=true]{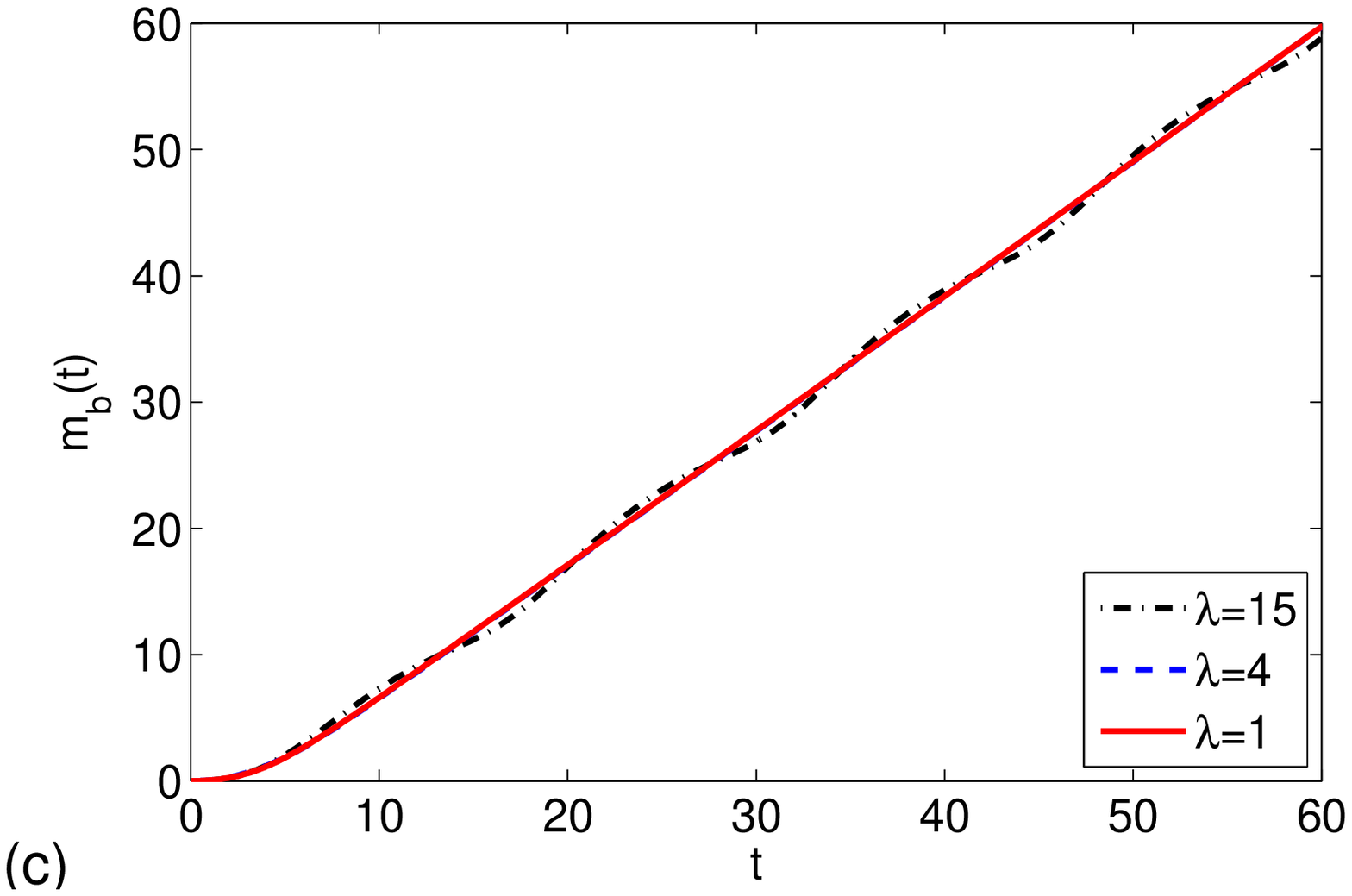} 
\includegraphics[angle=0,scale=0.4,keepaspectratio=true]{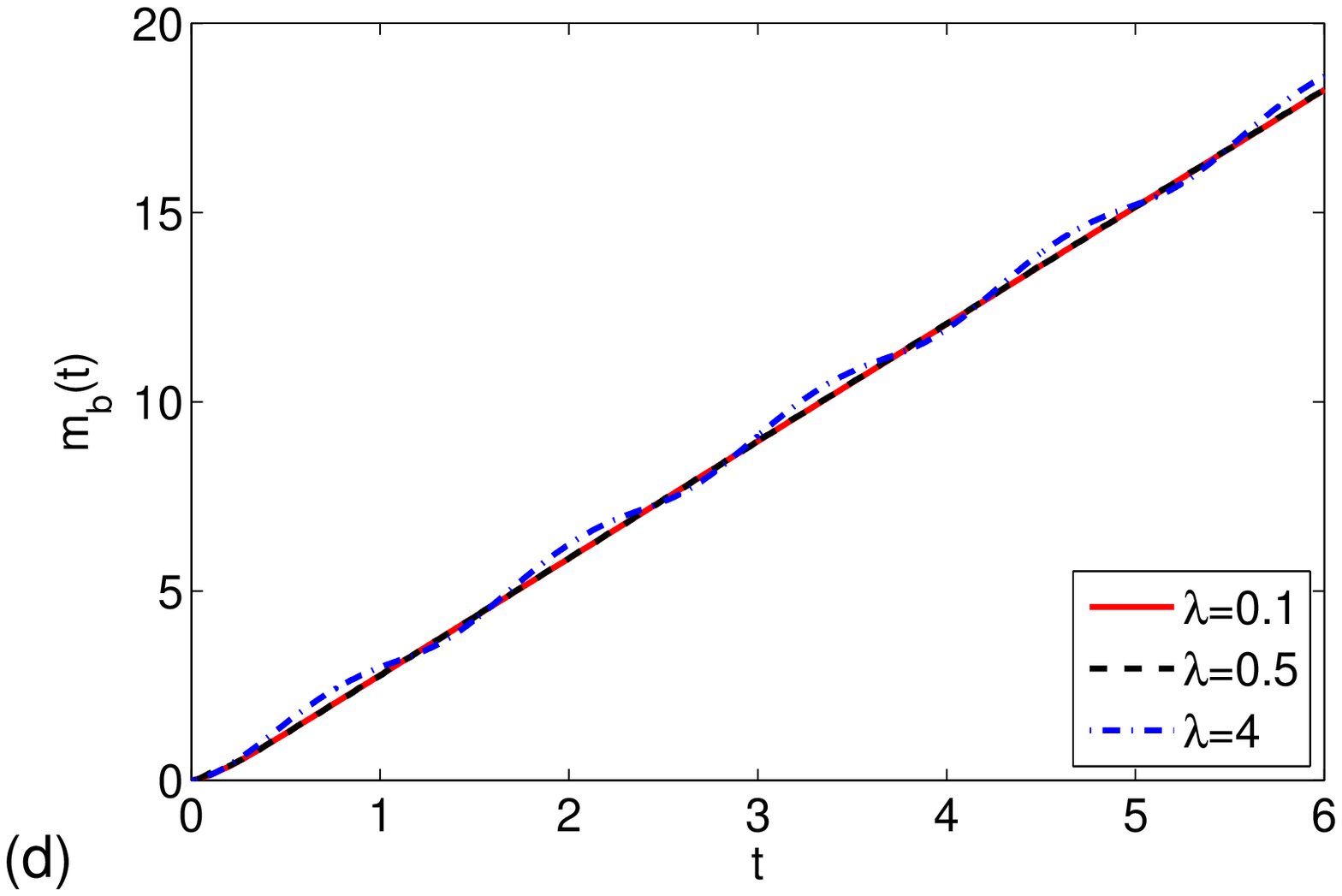} \\
\caption{(Color online) Front speed and burnt mass ($m_b(t)=\int_0^{t} v_f(t)~dt$) as function of time for a fixed $Pe=100$, two different Damk\"{o}hler
 ($Da=1$ in panel: a,c ; $Da=100$ in panel: b,d) and different $\lambda$.  }
\label{speedlambda}
\end{figure}

Then we have carried out simulations in which the compressibility is fixed ($\epsilon=0.5$) and we choose $\lambda$ approximately 
greater, lower or between the two thicknesses computed in the case of zero compressibility.\\
In Figure \ref{fronti} we show how the { geometrical} aspect of the reactive front changes by varying $\lambda$.
In the low density zones, the front thickness appears broader than in high density zones due to the decreasing of the local
Peclet and Damk\"{o}hler number ($Pe_l=\rho Pe$, $Da_l=\rho Da$).
Compressibility perturbation wrinkles the front in the small-wavelength limit,
whereas for large wavelengths it is only corrugated, since the entire reactive-diffusive front is embedded in a wavelength.
Nevertheless, even though the front does not appear stationary (even in a co-moving reference system) and it is noticeably distorted by the presence of compressibility,
the mean velocity of propagation ($v$) does not change, see Figure \ref{speedlambda}.
The wavelength of the perturbation controls the frequency and the magnitude
 of the instantaneous value of the front speed but does not affect the mean value. Since asymptotic propagation is not 
affected by  $\lambda$, from now on in all simulations we set $\lambda=1$. 

\begin{figure}[ht!]
\includegraphics[angle=0,scale=0.4,keepaspectratio=true]{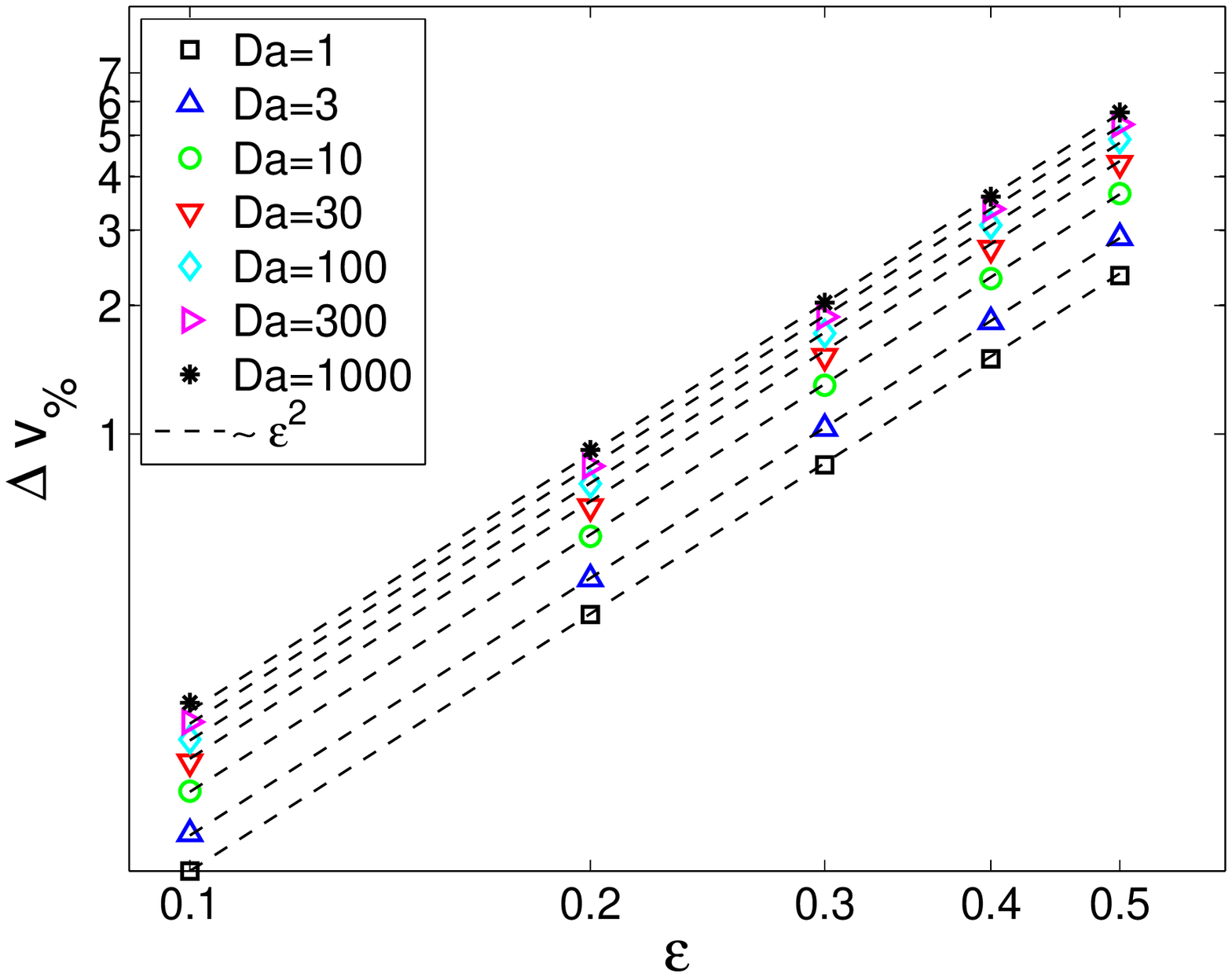} 
\includegraphics[angle=0,scale=0.4,keepaspectratio=true]{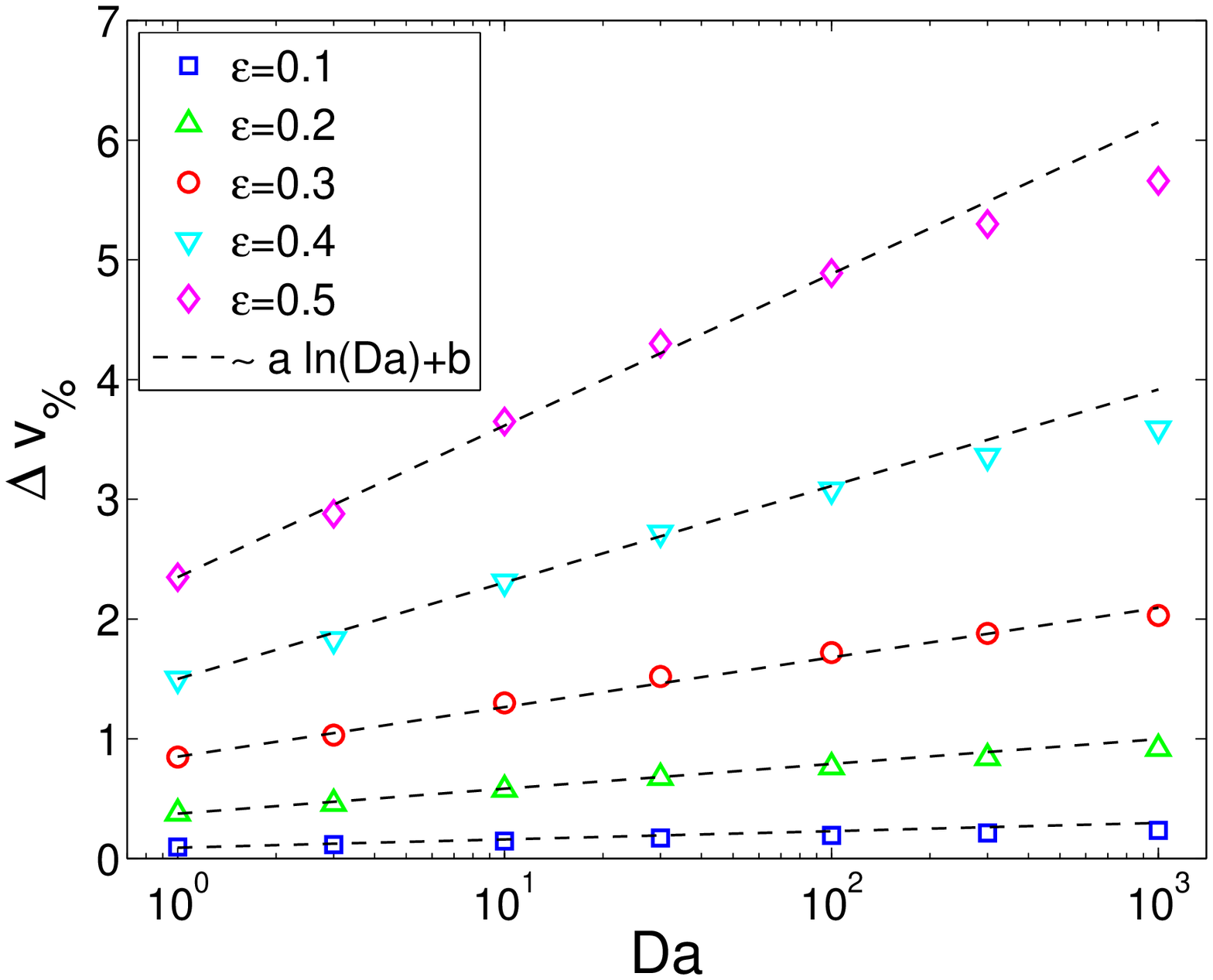} 
\caption{ (Color online) Comparison of the front speed between compressible and incompressible shear flow
 for a fixed Peclet number ($Pe=100$) at different compressibility magnitude $\epsilon$ (on the left) and for different Damk\"{o}hler (on the right).}
\label{mappe}
\end{figure}

In order to quantitatively characterize 
the effect of compressibility, we vary both the parameter $\epsilon$ and $Da$, with a fixed Peclet number. For this purpose, it is  convenient to define the percentage difference of the mean 
asymptotic front speed between the compressible and the incompressible case as follow:
\begin{equation}
\Delta v_\% = 100 ~\frac{v-v^0}{v^0}
\end{equation}
where $v^0$ is the asymptotic bulk burning rate, as defined in (\ref{vmedio}), for the incompressible case ($\epsilon=0$). Results are shown in Figure \ref{mappe}.\\
In general, in the regimes investigated here, we observe that the presence of compressibility can slightly improve the process of
reaction and the effects grow by increasing both $\epsilon$ and $Da$. 
For a fixed characteristic reaction rate (see Figure \ref{mappe}.a),
numerical simulations suggest a power (quadratic) law of the velocity enhancement as a function of the parameter $\epsilon$
\begin{equation}
 \Delta v_\%\sim \epsilon^2.
\end{equation}
Instead, the dependence on Damk\"{o}hler is much slower. 
As shown in Figure \ref{mappe}.b the parameter $\Delta v_\%$ is always
positive an it grows following (approximately) a logarithmic law:
\begin{equation}
 \Delta v_\% \sim a\ln(Da)+b
\end{equation}
where $a$ and $b$ may depend on $\epsilon$.
{ Therefore} even in the case of very strong compressibility ($\epsilon=0.5$) 
and very fast reaction ($\alpha=1000$) the difference never exceeds a modest $6\%$.
The effect of compressible wave perturbations appears therefore to be: 
i) wrinkling; ii) second-order enhancement.

\section{Compressible cellular flow}\label{sec:4}
We discuss now the case of cellular flows, i.e., 2D steady flows of amplitude $U_0$ composed by counter-rotating vortex
of dimension $L/2$. The compressibility is imposed in the following way:
\begin{align}
 \rho(x,y) &= \rho_0 C(x,y)  \\
\bar{u}(x,y) &= \left( \frac{ U_0  \sin \left(\frac{2 \pi y}{L}\right)\cos \left(\frac{2 \pi y}{L}\right) }{C(x,y)} , \frac{  -U_0 \cos \left(\frac{2 \pi y}{L}\right) \sin \left(\frac{2 \pi y}{L}\right) }{C(x,y)}\right)
\end{align}
We choose two different shapes for $C(x,y)$. In the first, 
that we call (a) case, the density of the mixture is 
higher in the centre of the vortex:\\
\begin{align}
C(x,y)=1+ \epsilon \left( \left| \sin\left(\frac{2 \pi x}{L}\right)\sin\left(\frac{2 \pi y}{L}\right) \right| -\frac{4}{\pi^2} \right)
\end{align}
In the second, that we call (b) case, the density 
is higher in the periphery of the vortex: 
\begin{align}
C(x,y)=1- \epsilon \left( \left| \sin\left(\frac{2 \pi x}{L}\right)\sin\left(\frac{2 \pi y}{L}\right) \right| -\frac{4}{\pi^2} \right)
\end{align}
\begin{figure}[ht!]
\includegraphics[angle=0,scale=0.5,keepaspectratio=true]{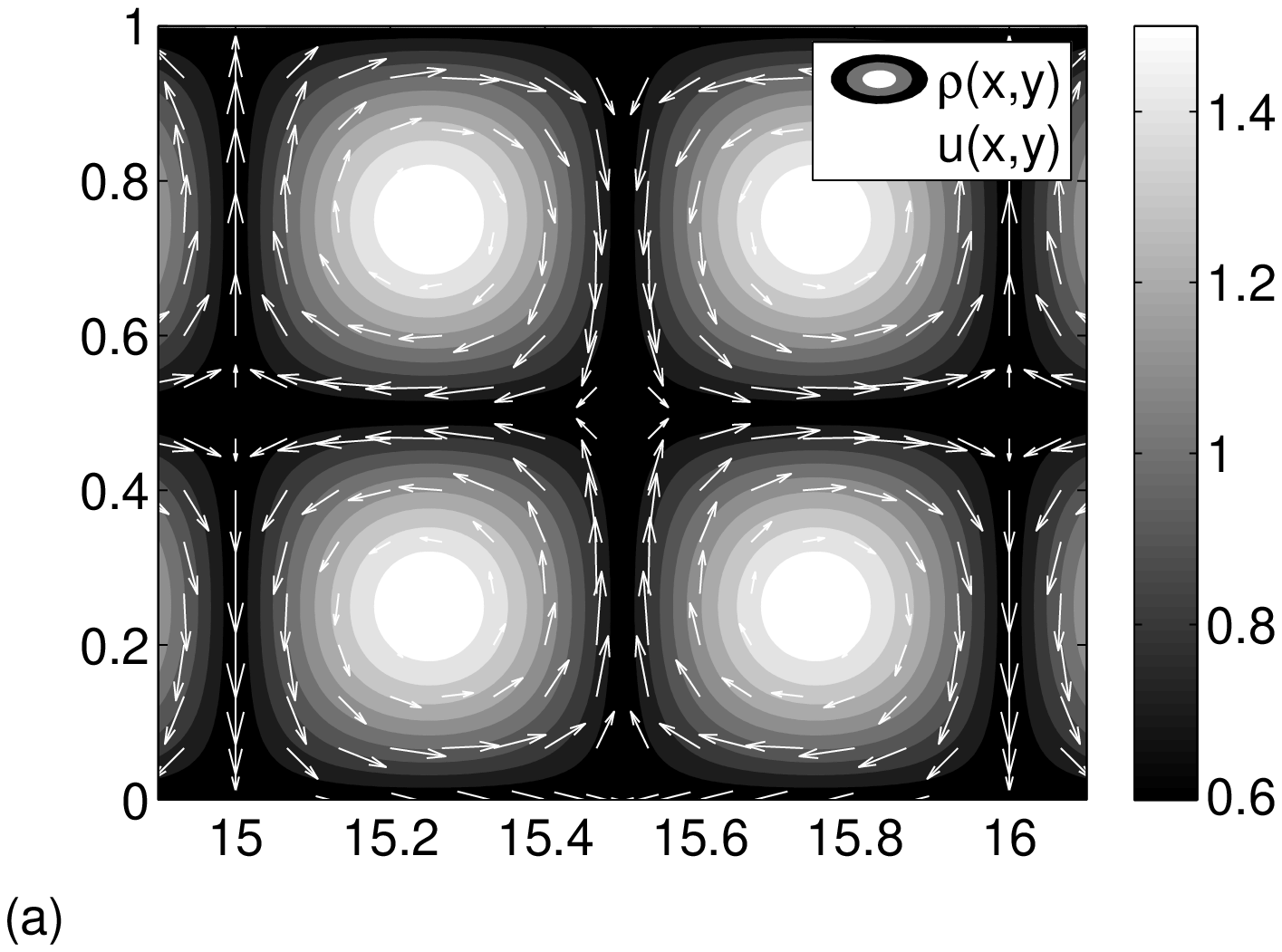} 
\includegraphics[angle=0,scale=0.5,keepaspectratio=true]{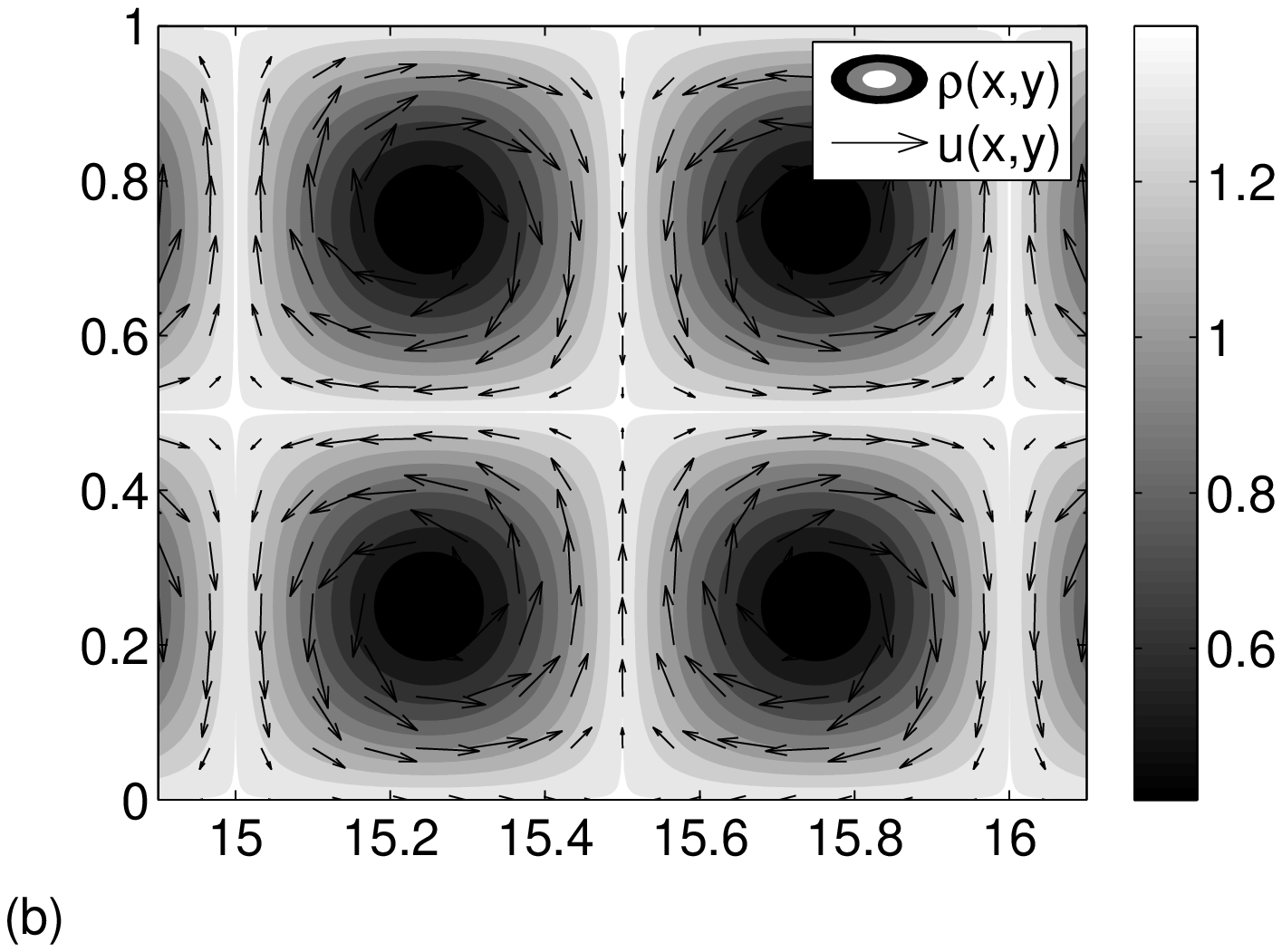} 
\caption{Compressible Cellular flow : (a) the density is higher in the centre of the vortex. (b) The density is higher in the outer region.}
\label{fig:cellular}
\end{figure}    
The two different configurations are shown in Figure~\ref{fig:cellular}. 
The constant factor $\frac{4}{\pi^2}$ has been introduced in order 
to have a density perturbation which is zero in average.  
As in the case of the shear flow, we study the dependence of
the dynamics on the compressibility intensity $\epsilon$ and
Damk\"{o}hler number in the more realistic case of fixed high Peclet
number.

\begin{figure}[ht!]
\includegraphics[angle=0,scale=0.45,keepaspectratio=true]{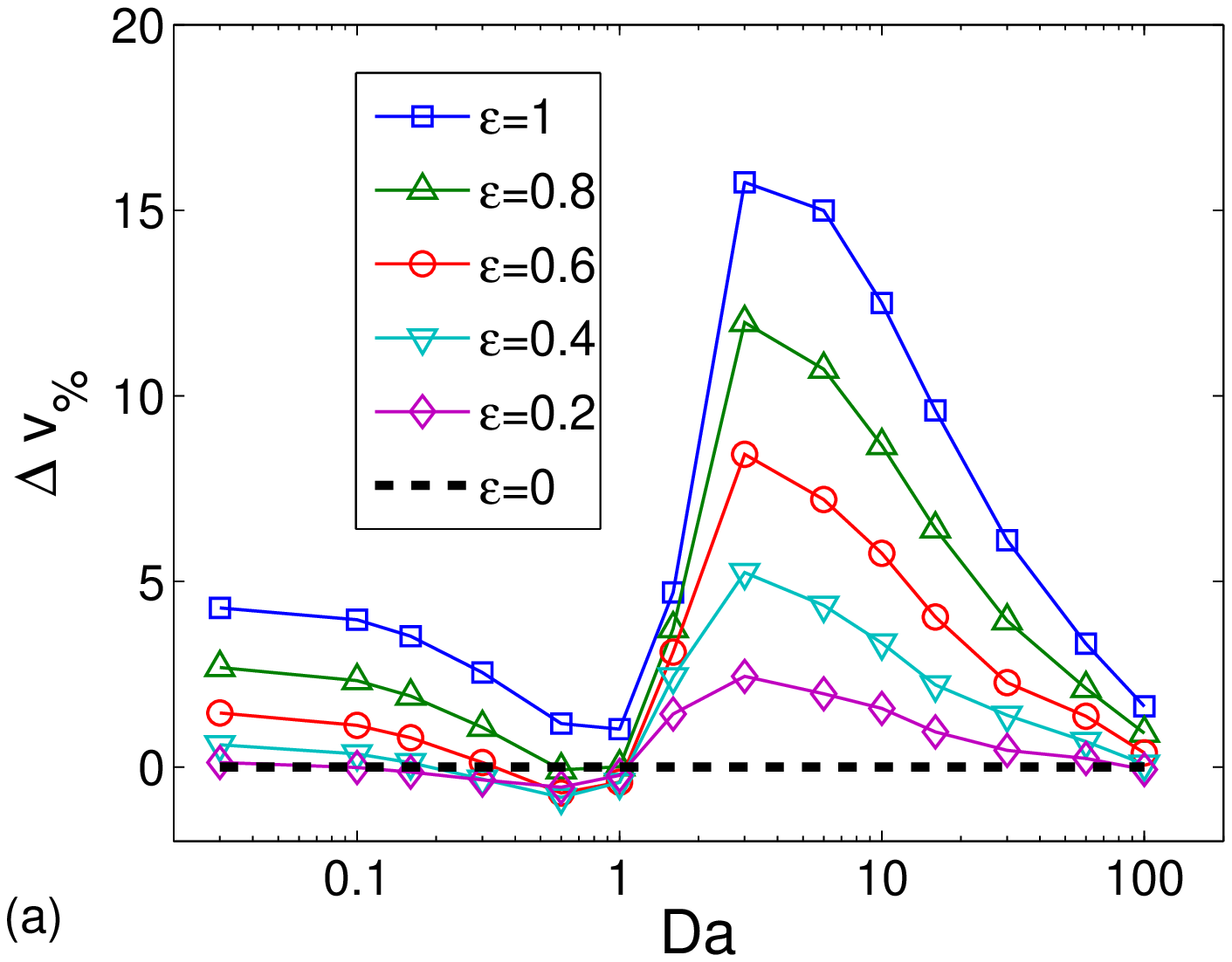} 
\includegraphics[angle=0,scale=0.45,keepaspectratio=true]{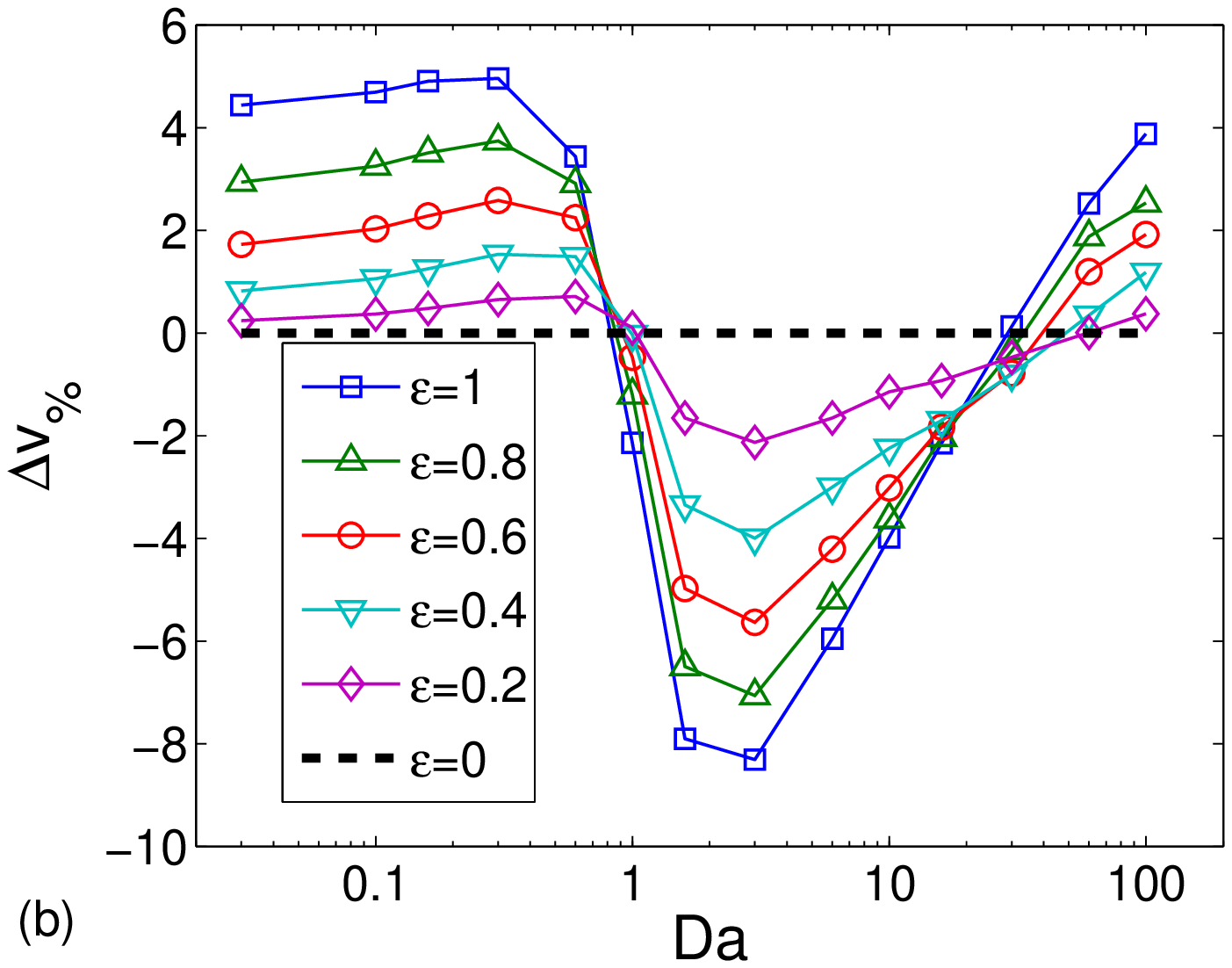} \\
\caption{(Color online) (a,b) Percentage difference between compressible and incompressible test case of the mean asymptotic bulk burning rate. $1/Pe=0.003$.
 In panel $(a)$ the density of the fluid is higher in the centre of the vortexes while in panel $(b)$ the density is higher
in the periphery.}
\label{fig:cellular2}
\end{figure}

We will consider a wider range of Damk\"{o}hler exploring the regimes
at $Da\ll1$, $Da\approx 1$ and $Da\gg1$.  Nevertheless we will remain
in regimes $PeDa>1$ which means that the characteristic diffusion time
is always larger than the reaction time. Unlike the shear flow, in the
cellular flow $\Delta v_\%$ does not show a monotonic dependence
neither on $\epsilon$ nor on $Da$, as it can be seen from Figure
\ref{fig:cellular2}.

Such a feature has been observed also for other configurations of
$C(x,y)$ (simulations not shown here) confirming that the non
monotonic behaviour of $\Delta v_\%$ is related to the whirling
geometry of the flow rather than to the choice of the density
perturbation. In the slow reaction regime ($Da \ll 1 $) it can be seen
in Fig. \ref{fig:cellular2} an almost Damk\"{o}hler-indipendent behaviour
of $\Delta v_\%$ in both cases (a) and (b). On the other hand, in a
middle range of Damk\"{o}hler where the combined effect of advection
and reaction is more intriguing, the two flow configurations show
opposite trends for $\Delta v_\%$, the reaction is faster, when the
density is higher in the centre (case (a)), whereas it is slower when
the perturbation is at the periphery (case (b)). Such a behaviour is
not surprising, since the interplay between reaction and diffusion in
the presence of closed streamlines can lead to a non trivial behaviour
also in the case of uncompressible flow~\cite{acvv1}, and the presence
of variations in the density of the flow can act in a very non
intuitive way.

\section{Conclusions}\label{sec:5}
We have studied the propagation of fronts through an
advection-diffusion-reaction equation where the nonlinear reaction
term is given by the classical FKPP source term. The advective flow is
generated by an imposed field which is perturbed by compressible
waves. The compressibility is controlled by the parameter
$\epsilon$. Two velocity fields have been considered: a shear-flow and
a cellular one.

In the considered flows, the front can be strongly affected by
compressibility and the compressibility field forces a strong
localization of density, but the quantitative differences with 
respect to the incompressible model appear modest (of the order 
of some percent). On the basis of previous studies, we do not
think that the presence of chaos (turbulent fluctuations) should
change much the scenario~\cite{ctvv}.

Some comments are in order to discuss the apparent difference
of the behaviour of $\Delta v_\%$ in the cases of shear flow
and cellular flow (see Figs.~\ref{mappe} and \ref{fig:cellular2}). 
The stream lines in the two cases are very different:
namely open and closed, respectively. In the shear flow,
the effect of compressibility on the front propagation
is only slightly modified with respect to the uncompressible 
case, since the front is mainly driven by the stream. On the
other hand, closed streamlines trigger entangled mechanisms
between reaction and diffusion, that, coupled with the 
compressibility generate highly non trivial features.
An example of this complicated behaviour can be found
in the non monotonic dependence of $\Delta v_\%$ 
by the Damk\"{o}hler number, or in the well apparent 
difference between cases (a) and (b) of the cellular flows
here considered.

Finally, it is interesting to note that a similar model has been
recently used for the study of population dynamics in turbulent
flows~\cite{benzi}:
\begin{equation}
 \frac{\partial{C}}{\partial t} +  {\bf \nabla} \cdot ({\bf u} C)  =D_0 \nabla^2 C+ \mu C (1-C)
\label{Bianco_Federicoeq6}
\end{equation}
where the scalar $C({\bf x},t)$ is the concentration of a
population~\cite{benzi}, which is the equivalent of our $\rho \theta$
in Eq.~(\ref{scalar2}).  When ${\nabla} \cdot {\bf u} \ne 0$,
clustering of the population near compression regions 
$({\nabla}\cdot{\bf u} < 0)$ is observed.  
In those regions, the concentration can take values greater 
then one and reaction rate on Eq.~(\ref{Bianco_Federicoeq6}) 
can be negative, so that the scalar $C({\bf x},t)$ is not a fractional 
parameter.  Within this model,
authors linked changes in the overall carrying capacity of the ecosystem (i.e. the density of biological mass of the  system) to the compressibility and its effect of localisation. 
Yet, present results show that the change of the carrying capacity
is not due to density localisation, but rather to the choice of a 
different reaction term, 
which allows negative rate in high density zones. Indeed, 
in the present work we have a strong density localisation but our FKPP
model for a fractional parameter does not allow negative rate. 
The results is that the average carrying capacity does not change even in presence of compressible flows. 
The analysis of present results for the Lagrangian displacement of passive reactive tracers and irreversible reaction dynamics is ongoing. 

\section{Acknowledgements}
We would like to thank Dr Guillaume Legros, P. Perlekar and Dr Roger Prud'homme for fruitful discussions.

\bibliographystyle{unsrt}
\bibliography{biblio}

\end{document}